\documentclass[aps,prd,superscriptaddress,nofootinbib,amsmath,amsfonts,preprintnumbers,notitlepage,10pt,english]{revtex4}
\usepackage{amsmath}
\usepackage{amssymb}
\usepackage{babel}
\usepackage{graphicx}
\usepackage{color}
\def\d{\delta}
\newcommand{\f}[2]{\frac{#1}{#2}}
\newcommand{\mk}[1]{\left( #1 \right)}
\newcommand{\kk}[1]{\left[ #1 \right]}
\newcommand{\ck}[1]{\left\{ #1 \right\}}
\newcommand{\be}{\begin{equation}}
\newcommand{\ee}{\end{equation}}
\newcommand{\bea}{\begin{eqnarray}}
\newcommand{\eea}{\end{eqnarray}}
\def\Mpl{M_{\rm Pl}}
\def\ini{{\rm ini}}
\def\erf{{\rm erf}}
\def\max{{\rm max}}



\begin{document}

\preprint{RESCEU-8/12}

\title{Reheating after $f(R)$ inflation}

\author{Hayato Motohashi}
\affiliation{Department of Physics, Graduate School of Science,
The University of Tokyo, Tokyo 113-0033, Japan}
\affiliation{Research Center for the Early Universe (RESCEU),
Graduate School of Science, The University of Tokyo, Tokyo 113-0033, Japan}

\author{Atsushi Nishizawa}
\affiliation{Yukawa Institute for Theoretical Physics, 
Kyoto University, Kyoto 606-8502, Japan}

\begin{abstract}%%%%%%%%%%%%%%%%%%%%%%%%%%%%%%%%%%%%%%%%%
The reheating dynamics after the inflation induced by $R^2$-corrected $f(R)$ model is considered. To avoid the complexity of solving the fourth order equations, we analyze the inflationary and reheating dynamics in the Einstein frame and its analytical solutions are derived. We also perform numerical calculation including the backreaction from the particle creation and compare the results with the analytical solutions. Based on the results, observational constraints on the model are discussed.
\end{abstract}

\date{\today}

\maketitle

\section{Introduction}%%%%%%%%%%%%%%%%%%%%%%%%%%%%%%%%%%%%%%%%%

Seeking the physical origin of two accelerated expansion regimes of the Universe, namely, the primordial inflation and the present cosmic acceleration, is one of the most important theoretical challenges of cosmology today. 
These unknown physical origins are referred by the primordial dark energy (DE) and the present DE. Various theoretical models have been proposed to accelerate the cosmological expansion.

Among those theoretical models, $f(R)$ gravity is a simple and nontrivial generalization of General Relativity. For a recent review, see Refs.~\cite{Sotiriou:2008rp,DeFelice:2010aj}. It introduces a function $f(R)$ in the action, where $R$ is Ricci curvature. This additional degree of freedom plays a role of a scalar field, which is called scalaron, and it can cause the accelerated expansion of the Universe. The original idea recognized as $R^2$ inflation model was proposed in Ref.~\cite{Starobinsky:1980te}, where de Sitter expansion was derived as a solution for the Einstein equation with quantum one-loop correction. After the accelerating expansion, the particles are gravitationally created through the oscillation of the scalaron, and it leads radiation dominated Universe~\cite{Zeldovich:1977,Starobinsky:1981vz,Vilenkin:1985md,Mijic:1986iv,Ford:1987}. The $R^2$ model predicts an almost scale-invariant spectrum, whose scalar and tensor components are consistent with recent observational data (See {\it e.g.}, \cite{Komatsu:2009}).

Later, the $R^2$ gravity was extended to a general function of $R$, namely $f(R)$, to describe the late time cosmic acceleration. After some early challenges, the viable $f(R)$ models were proposed~\cite{Hu:2007nk,Appleby:2007vb,Starobinsky:2007hu}, which can realize stable matter-dominated regime and subsequent late time acceleration. In these models, the expansion history of the Universe is close to that in the ${\rm \Lambda}$-cold dark matter (${\rm \Lambda}$CDM) model. From the model selection point of view, the key to distinguish the models is focusing on small deviation from the ${\rm \Lambda}$CDM model. As for background quantity, the equation of state parameter $w_{\rm DE}$ well characterizes the models. While it remains constant $w_{\rm DE}=-1$ in the ${\rm \Lambda}$CDM model, it is time dependent in the $f(R)$ gravity and even crosses the phantom divide at redshift $z\sim O(1)$~\cite{Hu:2007nk,Motohashi:2010tb,Motohashi:2011wy}. On the other hand, the growth of the matter density fluctuations is also useful to measure the deviation. Since in $f(R)$ gravity the effective gravitational constant depends on time and scale, the matter power spectrum is enhanced~\cite{Zhang:2005vt,Tsujikawa:2007gd,Hu:2007nk,Starobinsky:2007hu,Gannouji:2008wt,Tsujikawa:2009ku,Motohashi:2009qn,Narikawa:2009ux}. This enhancement not only measure the deviation but also provides another interesting consequence: $f(R)$ gravity admits larger neutrino mass. This is because massive neutrinos suppress evolution of matter fluctuations by free streaming, which cancels the enhancement in $f(R)$ gravity. As a result, $f(R)$ gravity allows larger neutrino mass up to $\sim 0.5$~eV~\cite{Motohashi:2010sj}. The constraint on sterile neutrino mass is also relaxed up to $\sim 1$~eV, which is consistent with recent experiments~\cite{Motohashi:2012wc}. Other distinguishable features of $f(R)$ gravity would be imprinted on cosmological gravitational waves \cite{Ananda:2008,Capozziello:2008,Alves:2009}. Future pulsar timing experiments and gravitational-wave detectors will be able to probe them directly and test gravity theories \cite{Lee:2008,Chamberlin:2011,Nishizawa:2009,Nishizawa:2010}.

However, the above viable $f(R)$ models still have theoretical problems~\cite{Starobinsky:2007hu}. If we start from some initial condition and calculate back to the past, the scalaron mass diverges quickly and the scalaron oscillates rapidly~\cite{Tsujikawa:2007xu}. Another problem is that the Ricci scalar also diverges at the past even if we include nonlinear effect~\cite{Appleby:2008tv}. This curvature singularity was also pointed out in Refs.~\cite{Frolov:2008uf,Kobayashi:2008tq}.
To solve these problems, $R^2$-corrected $f(R)$ model has been proposed~\cite{Appleby:2009uf}. This model is constructed from the late time acceleration part and $R^2$ term. Consequently, the reheating followed by inflation in this model is different from that in $R^2$ model, {\it i.e.}, the scalaron does not oscillate harmonically. 
Of course, we can construct the other specific functional forms that avoid singularities and describe both the primordial DE and the present DE. However, these functions belong to the same class and their behavior are similar~\cite{Motohashi:2012}. In this sense, it is worth to study one specific model in detail as an example of such a class of extended $f(R)$ models.

The aim of this paper is to investigate the evolution during the inflation and reheating regimes of the $R^2$-corrected $f(R)$ model in detail.
In the previous work~\cite{Appleby:2009uf}, the reheating dynamics has been analyzed in the Jordan frame. However, the field equation is fourth order differential equation, which makes the physical interpretation unclear. In addition, in their analysis, the inflation and the reheating are separately solved with different initial conditions. Therefore, we focus on the evolution of the scalaron rolling on a potential in the Einstein frame, which clarifies the physical picture and allows us to understand the dynamics intuitively. We start from a certain initial condition imposed during inflationary regime and numerically solve the transition from the inflation to the reheating and the following reheating dynamics. Thus, our analysis is more accurate than the previous work \cite{Appleby:2009uf} and is complementary to the analysis in the Jordan frame. 

This paper is organized as follows. In Sec.~\ref{sec-be}, we review the basic equations in $f(R)$ gravity in both the Jordan frame and the Einstein frame. We present the Einstein frame potential in the $R^2$-corrected $f(R)$ model and consider its characteristics analytically. In Sec.~\ref{sec-an}, we derive the analytic solutions of the inflation and reheating in the Einstein frame. We shall adopt the slow roll and the fast roll approximations and solve the field equations in each era. Sec.~\ref{sec-nr} contains the results of the numerical calculation. We confirm that the analytic solutions are sufficiently in agreement with the numerical results. We also consider the behavior at the end of the reheating. In Sec.~\ref{sec-rtcn}, we consider the connection between observables and model parameters and discuss its allowed ranges. Sec.~\ref{sec-cn} is devoted to conclusions and discussion. Throughout the paper, we adopt units $c=\hbar=1$.

\section{Basic equations}%%%%%%%%%%%%%%%%%%%%%%%%%%%%%%%%%%%%%%%%%
\label{sec-be}

We start to review the basic equations of $f(R)$ gravity in both the Jordan frame and the Einstein frame. To avoid confusion of the frames, we fix the subscript $J$ and $E$ to physical quantities in the Jordan frame and the Einstein frame, respectively. Otherwise we declare the frame in which the quantity is defined. $f(R)$ gravity is defined by the action
\be S=\int d^4x \sqrt{-g_J} \left[ \f{M_{\rm{Pl}}^2}{2} f(R_J) + {\cal L}_M(g^J_{\mu\nu}) \right] \;, \ee
where ${\cal L}_M$ is the Lagrangian density for the matter sector and $M_{\rm{Pl}}$ is the reduced Planck mass. By varying the action, we obtain the field equation in the Jordan frame
\begin{equation}
R^J_{\mu\nu} F(R_J) -\frac{1}{2} g^J_{\mu\nu} f(R_J) + ( g^J_{\mu \nu} \Box -\nabla_{\mu} \nabla_{\nu} ) F(R_J)
= \frac{T^J_{\mu\nu}}{M_{\rm{Pl}}^2} \;, 
\label{eq10}
\end{equation}
with 
\begin{equation}
F(R_J) \equiv \frac{d\,f(R_J)}{dR_J} \;, \quad \quad T_{\mu \nu}^J \equiv -\frac{2}{\sqrt{-g_J}} \frac{\delta (\sqrt{-g_J} {\cal{L}}_{\rm{M}})}{\delta g_J^{\mu \nu}} \;. \label{defF}
\end{equation}

We regard the Jordan frame as the physical frame. However, for our purpose analyzing the inflation and the reheating in $f(R)$ gravity, the formulation in the Einstein frame is useful because it contains the additional degree of freedom more explicitly as a scalar field and enables us to use the analogy of the single-field inflation. We can recast the theory to the Einstein gravity with a scalar field by choosing the conformal transformation of the metric as $g^E_{\mu\nu}\equiv F(R_J) g^J_{\mu\nu}$. The canonical scalar field $\phi$ is defined by 
\be F(R_J)\equiv e^{\sqrt{\f{2}{3}}\f{\phi}{\Mpl}}. \label{FRJ} \ee 
By the conformal transformation, the action is rewritten as
\be S=\int d^4x \sqrt{-g_E} \kk{\f{M_{\rm{Pl}}^2}{2} R_E-\f{1}{2}g_E^{\mu\nu} \partial_\mu \phi \partial_\nu \phi-V(\phi) +{\cal L}_M \left(e^{-\sqrt{\f{2}{3}}\f{\phi}{\Mpl}} g^E_{\mu\nu} \right)}, \ee
with the potential term
\be V(\phi)=\frac{\Mpl^2}{2} \f{R_J(\phi)F(R_J(\phi))-f(R_J(\phi))}{F(R_J(\phi))^2}. \label{EFpot} \ee
Then, the Einstein equation in the Einstein frame reduces to 
\bea
H_E^2&=&\f{1}{3\Mpl^2}\kk{\f{1}{2}\mk{\f{d\phi}{dt_E}}^2 + V(\phi) + \rho_E}, \label{Eeq1} \\
\f{dH_E}{dt_E}&=&-\f{1}{2\Mpl^2}\kk{\mk{\f{d\phi}{dt_E}}^2 + \rho_E+P_E}. \label{Eeq2}
\eea
The equation of motion for the scalar field is 
\be \f{d^2\phi}{dt_E^2}+3H_E\f{d\phi}{dt_E}+V_{,\phi}(\phi)=\f{1}{\sqrt{6} \Mpl} (\rho_E-3P_E). \label{Eeq3} \ee

From the conformal transformation, the time and scale factor in both frames are connected by
\be dt_J=e^{-\f{1}{\sqrt{6}}\f{\phi}{\Mpl}}dt_E, \quad a_J=e^{-\f{1}{\sqrt{6}}\f{\phi}{\Mpl}}a_E. \label{ta} \ee
The transformation of the Hubble parameter is derived from the above definitions,
\be H_J=e^{\f{1}{\sqrt{6}}\f{\phi}{\Mpl}}\mk{H_E-\f{1}{\sqrt{6}\Mpl}\f{d\phi}{dt_E}}. \label{HJ} \ee
By definition in Eq.~\eqref{defF}, the energy momentum tensors of the matter sector in both frames are connected by
\be T^E_{\mu\nu}= e^{-\sqrt{\f{2}{3}}\f{\phi}{\Mpl}} T^J_{\mu\nu}. \ee
For perfect fluid $T^{\mu}_{\nu}={\rm diag}(-\rho, P,P,P)$ in each frame, the energy density and the pressure are related as
\be \rho_E=e^{-2\sqrt{\f{2}{3}}\f{\phi}{\Mpl}}\rho_J,\quad P_E=e^{-2\sqrt{\f{2}{3}}\f{\phi}{\Mpl}}P_J. \ee
Note that the energy density in the Einstein frame couples with the scalaron.

In the inflation and reheating in $f(R)$ gravity, there is no inflaton field from the point of view in the Jordan frame. Consequently, particle creation occurs not through the decay of the inflaton but through the gravitational reheating~\cite{Zeldovich:1977,Starobinsky:1981vz,Vilenkin:1985md,Mijic:1986iv,Ford:1987}. Let us consider the gravitational particle creation in the Jordan frame. We introduce a minimally or nonminimally coupled massless scalar field $\chi$, which describes the created particles, into the matter action,
\be S=\int d^4x \sqrt{-g_J} \kk{\f{\Mpl^2}{2}f(R_J) -\f{1}{2}g_J^{\mu\nu} \partial_{\mu} \chi \partial_{\nu} \chi -\f{1}{2} \xi R\chi^2}. \ee
Since the scalar field $\chi$ couples with the metric in the Jordan frame, the radiation (massless scalar particle) is created purely gravitationally.
Adopting the standard treatment of the quantum field theory in curved spacetime, we can expand $\chi$ in Fourier modes with the annihilation and creation operators. Then, computing the Bogolubov coefficients in the expanding Universe, we obtain the number density of the created scalar particles~\cite{Zeldovich:1977,Starobinsky:1981vz,Vilenkin:1985md,Mijic:1986iv},
\be n_J(t_J)=\f{(1-6\xi)^2}{576\pi a_J^3} \int_{-\infty}^{t_J} dt'_J a_J^3R_J^2. \ee
The above equation holds regardless of the functional form of $f(R)$. Note that the particle creation is sourced by Ricci curvature. Since the Ricci curvature is significantly suppressed during the reheating era as we shall see below, particle creation hardly occurs. On the other hand, inflationary dynamics is the same as that of the $R^2$ inflation. Therefore, we can use the approximated formula for the $R^2$ model and turn off the particle creation during the reheating era when we perform numerical calculation. In the $R^2$ model with $f(R_J)=R_J+R_J^2/(6M^2)$, the energy density of the created particles is 
\be \rho_J(t_J)=\f{g_* M (1-6\xi)^2}{1152\pi a_J^4} \int_{-\infty}^{t_J} dt'_J a_J^4 R_J^2 \label{rhoint}, \ee
where $g_*$ denotes the relativistic degree of freedom. In the present paper, we consider minimally coupled massless scalar field and set $\xi=0$ hereafter.
The evolution equation for the energy density of radiation is then
\be \f{d\rho_J}{dt_J}=-4H_J\rho_J+\f{g_*MR_J^2}{1152\pi}\;. \label{Jeq1} \ee 
The pressure is obtained from the energy conservation equation,
\be P_J=\f{\rho_J}{3}-\f{g_*MR_J^2}{3456\pi H_J}. \ee

Finally, we introduce $gR^2$-AB model~\cite{Appleby:2009uf}, which describes the accelerated expansions in both the early and the present Universes,
\be f (R_J)=(1-g)R_J+g M^2\d \log\kk{\f{\cosh(R_J/\d M^2-b)}{\cosh b}}+\f{R_J^2}{6M^2}, \label{gR2AB} \ee
where $g,~b,~\d$, and $M$ are positively-defined model parameters. 
$\d$ describes the ratio between the energy scale of the present DE to the primordial DE and takes dramatically small value.
$M$ will be fixed  to $M/\Mpl \approx 1.2\times 10^{-5}$ later in Sec.~\ref{sec-rtcn} by the temperature fluctuations of the cosmic microwave background (CMB) anisotropy. $g$ is constrained in the range of $0<g<1/2$ by the stability conditions of $f(R)$ gravity: $F(R_J)>0$ and $dF(R_J)/dR_J>0$. Moreover, $g$ and $b$ are further constrained by the other stability condition as we shall show in Sec.~\ref{sec-rtcn}.
The function $F(R_J)$ is given by 
\be F(R_J)=1-g+\f{R_J}{3M^2}+g \tanh (R_J/M^2\d -b). \label{FgABR2} \ee
The above $f(R_J)$ function is equivalently written in the following form:
\begin{align}
f(R_J) &=R_J-\frac{R_{\rm{vac}}}{2} + g M^2 \delta \log \left[ 1+e^{-2(R_J/M^2\delta -b)} \right]+\frac{R_J^2}{6M^2} \;, 
\label{eq58} \\
R_{\rm{vac}} &\equiv 2 g M^2 \delta \left\{ b+\log(2\cosh b) \right\}. 
\end{align}
In Eq.~(\ref{eq58}), the fourth term dominates at high curvature regime $R_J \gg M^2$ and causes inflation~\cite{Starobinsky:1980te}. The third term alters the reheating dynamics after the inflation, which is characteristic of the $gR^2$-AB model. The second term plays the same role as the current cosmological constant. By substituting the action into Eq.~(\ref{eq10}), the equation of motion in the Jordan frame is given by
\begin{align}
&H_J^{''} H_J-\frac{1}{2} (H_J^{'})^2+3 H_J^{'} H_J^2 +\frac{1-g}{2} M^2 H_J^2 -\frac{g}{2} M^2 (H_J^{'}+H_J^2) \tanh \left[ \frac{R_J}{M^2\delta}-b \right] \nonumber \\
&+\frac{g}{12} M^4\delta \ln \left[ \frac{\cosh (R_J/M^2\delta -b)}{\cosh b} \right] + \frac{3g(H_J^{''} H_J + 4 H_J^{'} H_J^2)}{\delta \cosh^2 (R_J/M^2\delta -b)}= \frac{M^2}{6M_{\rm{Pl}}^2} \rho_J \;. 
\label{eq11}
\end{align}
The prime denote time derivative with respect to $t_J$. As expected from the specific form of $f(R_J)$ function, the equation of motion in the Jordan frame is nonlinear differential equation and quite complicated to solve.

\begin{figure}[t]
\centering
\includegraphics[width=85mm]{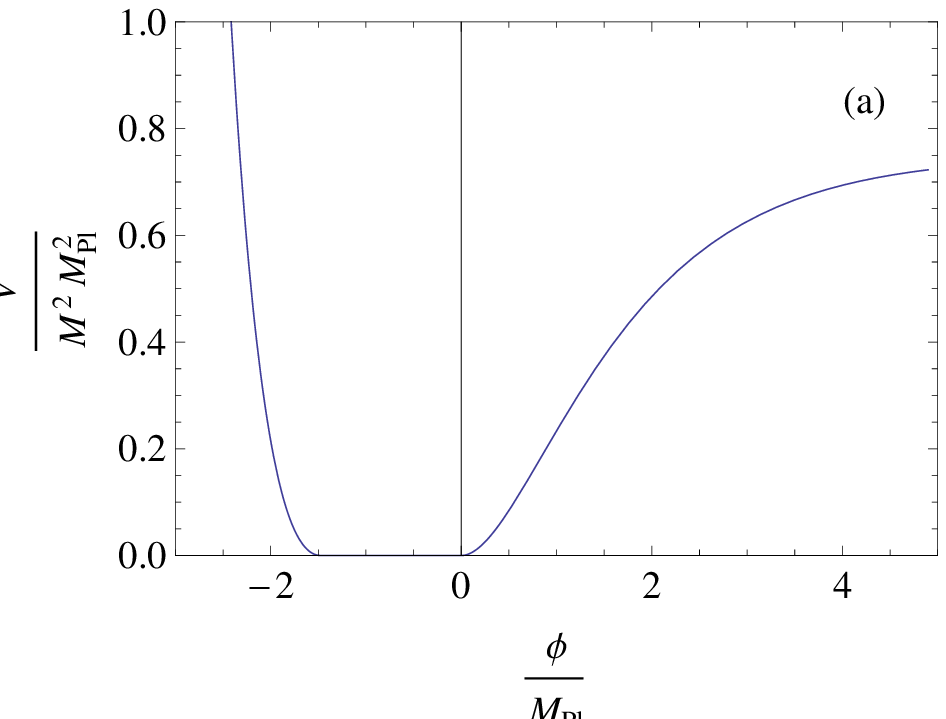}
\includegraphics[width=85mm]{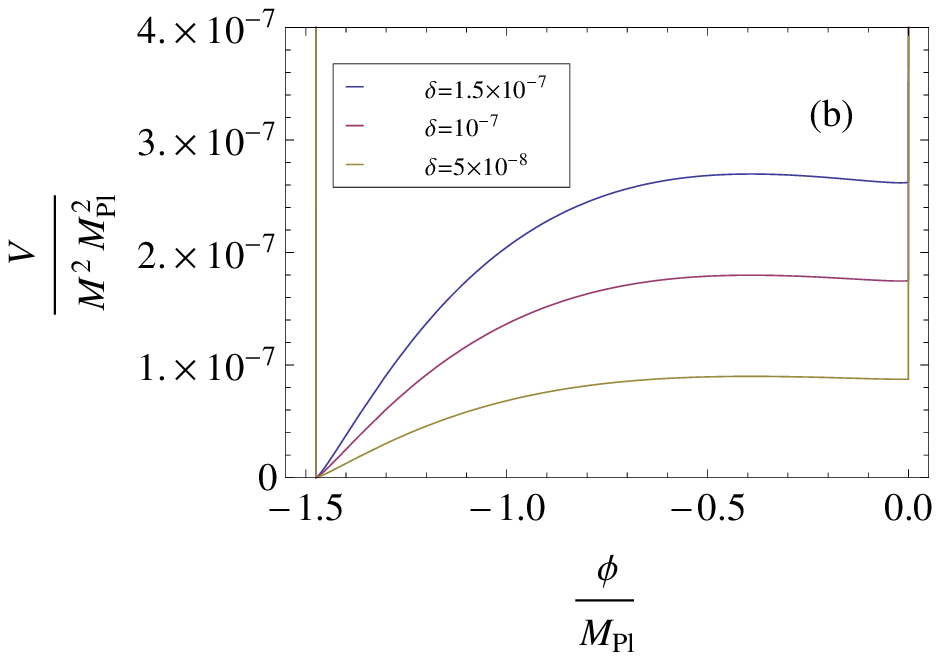}
\includegraphics[width=85mm]{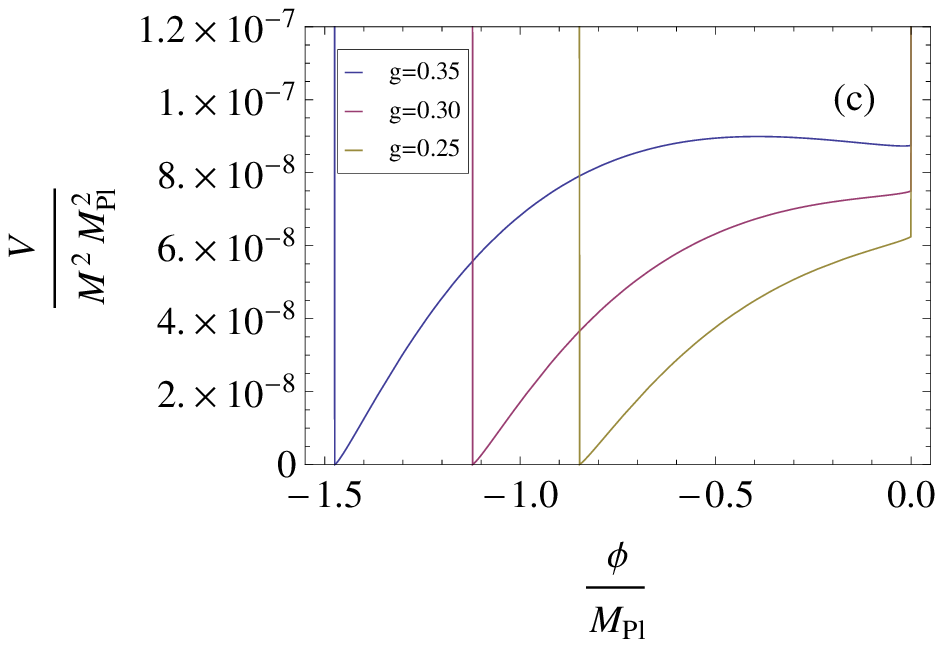}
\includegraphics[width=85mm]{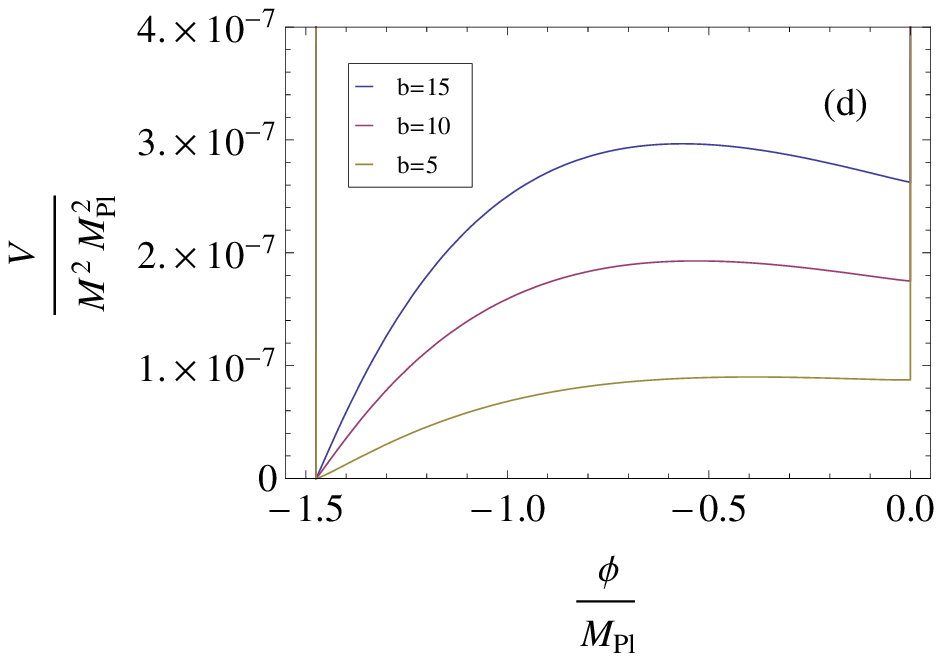}
\caption{
Inflaton potential of the $gR^2$-AB model in the Einstein frame. (a): Potential for $g=0.35, b=5, \d=5\times 10^{-8}$. For $\phi>0$ and $\phi<\sqrt{6}\Mpl\log \gamma$, the potential is similar to that in pure $R^2$ inflation. On the other hand, for $\sqrt{6}\Mpl\log \gamma<\phi<0$, there is the characteristic plateau. The scalaron starts slow rolling from $\phi>0$, and enters the plateau with large kinetic energy and oscillates inside it. (b) - (d): How plateau changes its shape when one parameter is changed from the value in (a). The typical height of the plateau is $bg\d$ from the analytic estimation. }
\label{fig:pot}
\end{figure}

The dynamics of inflation and reheating can be more intuitively understood from the potential in the Einstein frame, which is depicted in Fig.~\ref{fig:pot} (a). From the definition of Eq.~\eqref{EFpot}, we can interpret the shape of the potential in the following way. First, we notice that from Eq.~\eqref{FgABR2}, $F(R_J)$ is almost a step function at $R_J/M^2\simeq b\d$ with the change of the amplitude from $F=1-2g$ to $1$. In terms of the scalaron, the step corresponds from $\phi/\Mpl=\sqrt{6}\log\gamma$ to $0$, where $\gamma\equiv \sqrt{1-2g}$. In other words, during the scalaron moving in the range $\sqrt{6}\log\gamma \leq \phi/\Mpl \leq 0$, $R_J$ remains almost constant value, $R_J/M^2\simeq b\d$. Outside this interval, $F(R_J)$ is approximated by $F\simeq 1+R/3M^2$ for $R_J/M^2 > b\d$, {\it i.e.},  $\phi/\Mpl>0$, and $F\simeq 1-2g+R/3M^2$ for $R_J/M^2 < b\d$, {\it i.e.}, $\phi/\Mpl<\sqrt{6}\log\gamma$, respectively. By using these approximations, we can derive the potential analytically in terms of $\phi$:
\be
\f{V(\phi)}{\Mpl^2M^2}\simeq \left\{
\begin{array}{ll}
\displaystyle \f{3}{4}\mk{1-e^{-\sqrt{\f{2}{3}}\f{\phi}{\Mpl}}}^2, &\quad (\phi/\Mpl> 0) \\
\displaystyle \f{3}{4}\mk{1-\gamma^2 e^{-\sqrt{\f{2}{3}}\f{\phi}{\Mpl}}}^2. &\quad (\phi/\Mpl< \sqrt{6}\log \gamma)
\end{array}
\right. \label{potER2}
\ee
In these two regions, the potential is the same as that in the $R^2$ inflation. On the other hand, the characteristic plateau shows up for $\sqrt{6}\log\gamma<\phi/\Mpl<0$. By using $R_J/M^2\simeq b\d$, we can approximate the potential as
\be \f{V(\phi)}{\Mpl^2M^2}\simeq\f{b\d e^{\sqrt{\f{2}{3}}\f{\phi}{\Mpl}}-f(b\d M^2)/M^2}{2 e^{2\sqrt{\f{2}{3}}\f{\phi}{\Mpl}}}. \quad (\sqrt{6}\log \gamma<\phi/\Mpl< 0) \ee
From this expression, we can estimate the height of the bump in the plateau. We can obtain the position of the local maximum by solving $V'=0$. The solution is $\phi/\Mpl\simeq \sqrt{3/2}\log[2(1-2g)+b\d/3]\equiv \phi_m/\Mpl$. The potential has the local maximum when $\phi_m$ satisfies $\sqrt{6}\log \gamma < \phi_m/\Mpl < 0$, namely, $(3+b\d)/12 \lesssim g \lesssim (3+b\d)/6$. For $\d\ll 1$, this condition amounts to $1/4 \lesssim g \lesssim 1/2$. Therefore, in Fig.~\ref{fig:pot} (c), the potential for $g=0.35$ possesses the local maximum and the false vacuum. However, as $g$ approaches $g=1/4$, the local maximum is likely to disappear.
The potential heights at the right edge, the local maximum, and the left edge are given by
\be \f{V(0)}{\Mpl^2M^2}\simeq bg\d,\quad \f{V_\max}{\Mpl^2M^2}\simeq \f{b\d}{8(1-2g)} , \quad \f{V(\sqrt{6}\Mpl\log \gamma)}{\Mpl^2M^2}\simeq 0, \label{V0Vmax} \ee
where we used $\d\ll 1$ and $\log(\cosh b)\simeq b$. These estimations well explain the parameter dependences of the potential shape in Fig.~\ref{fig:pot} (b) - (d).

Next, let us consider the evolution of the scalaron.
The scalaron starts slow rolling from $\phi>0$ and plays a role of the inflaton. For $\phi>0$, the potential is the same as that of a pure $R^2$ model and is almost independent of model parameters $g,~b$ and $\d$. Thus, the scale factor in the Einstein frame experiences quasi-de-Sitter expansion. In this case, the scale factor in the Jordan frame also evolves exponentially, because the amplitude of $\phi$ slowly varies and thus the scale factors in both frames are related by multiplying an almost constant factor. As the scalaron approaches $\phi=0$, it rolls faster and enters the potential plateau with the kinetic energy larger than the potential energy. Then the scalaron oscillates in the plateau, gradually loses its kinetic energy, and finally reaches the false vacuum at $\phi=0$ because the chameleon effect lifts up the potential when the energy density of matter does not negligible~\cite{Motohashi:2012}. During this oscillation, the scale factor in the Jordan frame undergoes the periodic evolution due to the exponential factor in Eq.~(\ref{ta}). We shall see these situations in the next section.

%==================== Figure ====================
%==================== Figure ====================

\section{Analytic Solutions}%%%%%%%%%%%%%%%%%%%%%%%%%%%%%%%%%%%%%%%%%
\label{sec-an}

To investigate the reheating dynamics in the Jordan frame, we work in the Einstein frame and derive the analytical expressions for the motion of the scalaron in the inflaton potential. For $\phi>0$, the potential is almost reduced to that of a pure $R^2$ model, then we can use the slow-roll approximation in Sec.~\ref{ssec-sr} and derive the analytic solutions. After the end of the inflation, the kinetic energy of the scalaron is dominant. Therefore, we can neglect the small structure of the potential during the oscillation in the plateau and use the approximation, $R_J\sim b\d$. We shall explore this case in Sec.~\ref{ssec-fr}. Since in both the slow-roll and oscillation regimes the energy density of the created radiation is subdominant in comparison with the energy density of the inflaton, we neglect its backreaction to the background dynamics to derive the analytic solutions. However, the energy density of radiation eventually becomes the same order as the total energy of the inflaton, and the reheating ends. 
In the following, we define dimensionless variables $\hat t=Mt,~\hat\phi=\phi/\Mpl,~\hat V=V/M^2\Mpl^2,~\hat H=H/M,~\hat R=R/M^2,~\hat \rho_{\rm{r}}=\rho_{\rm{r}}/\Mpl^2M^2$ and abbreviate the hat in this section to avoid the complexity of notation.

\subsection{Slow roll approximation}%%%%%%%%%%%%%%%%%%%%%%%%%%%%%%%%%%%%%%%%%
\label{ssec-sr}

The inflaton starts to roll down the potential at $t_E=t_{E,\ini}$ ($t_J=t_{J,\ini}$) with small kinetic energy compared to the height of the potential. The energy density of the radiation is also negligible. The field equations \eqref{Eeq1} -- \eqref{Eeq3} with slow roll approximations are
\bea
H_E&=&\sqrt{\f{V(\phi)}{3}},\\
\dot H_E&=&-\f{V_\phi(\phi)^2}{6V},\\
\dot\phi&=&-\f{V_\phi(\phi)}{\sqrt{3V(\phi)}},
\eea
where the dot denotes the derivative with respect to the time in Einstein frame and the potential is approximated by Eq.~\eqref{potER2}.
We set the initial condition as $\phi=\phi_\ini$. Other initial quantities are fixed from the above equations with the approximated potential.
By substituting the potential into the field equations, we can derive the following solutions:
\bea
\phi(t_E)&=&\sqrt{\f{3}{2}}\log\kk{d_\ini^2-\f{2}{3}(t_E-t_{E,\ini})}, \label{phisr} \\
\dot \phi(t_E)&=&-\sqrt{\f{2}{3}}\kk{d_\ini^2-\f{2}{3}(t_E-t_{E,\ini})}^{-1}, \label{dphisr} \\
H_E(t_E)&=&\f{1}{2}\kk{1-\mk{d_\ini^2-\f{2}{3}(t_E-t_{E,\ini})}^{-1}},
\label{HEsr} \\
a_E(t_E)&=& a_{E,\ini} e^{(t_E-t_{E,\ini})/2} \kk{1-\f{2}{3}d_\ini^{-2}(t_E-t_{E,\ini})}^{3/4}, \label{aesr}\\
R_J(t_E)&=&3(e^{\sqrt{\f{2}{3}}\phi(t_E)}-1), \label{rjsr}
\eea
where we use $d_\ini\equiv e^{\f{\phi_\ini}{\sqrt{6}}}$ instead of $\phi_\ini$ itself.

The remaining task is to write down $t_E$ in terms of $t_J$ and to convert the above solutions.
Jordan frame time is derived by integrating Eq.~\eqref{ta}, 
\be t_J(t_E)=t_{J,\ini}-3\sqrt{d_\ini^2-\f{2}{3}(t_E-t_{E,\ini})}+3d_\ini, \ee
and its inverse function is
\be t_E(t_J)=t_{E,\ini}+(t_J-t_{J,\ini})\kk{d_\ini-\f{1}{6}(t_J-t_{J,\ini})}. \label{tetjsr} \ee
Thus, substituting Eq.~\eqref{tetjsr} into the solutions~\eqref{phisr} -- \eqref{rjsr} and using Eqs.~\eqref{ta} and \eqref{HJ},  we obtain the analytic solutions in terms of the Jordan frame quantities.

Next, we derive an analytic solution for $\rho_J$ by performing integration in Eq.~\eqref{rhoint} with Eqs.~\eqref{ta}, \eqref{phisr}, \eqref{aesr}, \eqref{rjsr} and \eqref{tetjsr},
\begin{align}
\rho_r(t_J)&=\f{g_*}{1152\pi}\mk{\f{M}{\Mpl}}^2
\f{1}{16S^2} \nonumber \\
& \times \kk{-2\sqrt{3}S(4S^4-14S^2+15)+3e^{S^2}\ck{-6d_\ini e^{-3d_\ini^2}(12d_\ini^4-14d_\ini^2+5)+5\sqrt{3\pi}\mk{\erf(\sqrt{3}d_\ini)+\erf(S)}}}, \label{rhoJsr}
\end{align}
where $S\equiv (t_J-t_{J,\ini}-3d_\ini)/\sqrt{3}$ and $\erf(x)$ is the error function. Hereafter, we refer $\rho_J$ as $\rho_r$, especially denoting radiation component.

Finally, we focus on boundary conditions. By using the above analytic solutions, we define the time $t_E=t_{E0}$ or $t_J=t_{J0}$ when the inflaton reaches $\phi=0$ for the first time. Strictly speaking, when $\phi\simeq 0$, the slow-roll approximation is not valid anymore and the boundary conditions are very sensitive due to the the sudden transition of the potential at $\phi\simeq 0$. Therefore, we should use the boundary conditions obtained from numerical computation. We only rely on the analytical boundary values as the estimator. We shall revisit this point in Sec.~\ref{sec-nr}.

The time $t_{E0}$ is analytically estimated by using the analytic solution \eqref{phisr},
\be t_{E0}=t_{E,\ini}+\f{3}{2}(d_\ini^2-1). \ee
In terms of Jordan frame time,
\be t_{J0}=t_{J,\ini}+3(d_\ini-1). \ee
The boundary conditions are 
\be
\phi_0=0, \quad 
\dot \phi_0=-\sqrt{\f{2}{3}}, \quad 
a_{E0}=a_{E,\ini}d_\ini^{-2}e^{\f{3}{4}(d_\ini^2-1)}.
\ee
From Eq.~\eqref{Eeq1},
\be
H_{E0}=\f{1}{3},\quad
H_{J0}=\f{2}{3}.\label{eq8}
\ee
The energy density $\rho_r$ at $t_J=t_{J0}$ is given by setting $S=-\sqrt{3}$ in Eq.~\eqref{rhoJsr} and taking large $\phi_{\rm{ini}}$ limit, {\it i.e.}, $d_\ini \gg 1$,
\begin{equation}
\rho_{r0} = \f{c_0 g_*}{1152\pi}\mk{\f{M}{\Mpl}}^2 \;. \nonumber 
\end{equation}
The coefficient $c_0$ is analytically given by $\kk{18+5e^3\sqrt{3\pi}\mk{1-\erf(\sqrt{3})}}/16 \approx 1.4$, but at this intermediate stage from the slow roll to the fast roll, the slow-roll approximation does not give the precise value. So we will use the result of the numerical calculation for the precise determination of $c_0$ later. We notice that since $\rho_{r0}\ll 3H_{J0}^2$ for typical values $g_*=106.75$ and $M \ll M_{\rm{Pl}}$, we can neglect the energy density of the created radiation for the purpose of deriving analytic formulas of reheating dynamics in the next subsection.

\subsection{Fast roll approximation}%%%%%%%%%%%%%%%%%%%%%%%%%%%%%%%%%%%%%%%%%
\label{ssec-fr}

When the inflaton falls down to the plateau on the bottom of the potential that lies $\sqrt{6}\log\gamma<\phi<0$, its kinetic energy is much greater than the potential energy, which is typically of the order of $bg\d$. Moreover, it is also greater than the energy density of radiation. Thus, we will use two approximations to derive the analytic solutions. First, the energy density created at the slow roll regime is not dominant as we mentioned above. Second, particle creation is negligible during the fast-roll regime because the source term $R_J^2$ keeps an almost constant value $b^2\d^2 \ll 1$. Thus, we can use the fast-roll approximation in Eqs.~(\ref{Eeq1}) -- (\ref{Eeq3}):
\bea
&&H_E^2=\f{\dot\phi^2}{6},\label{sr1}\\
&&\dot H_E=-\f{\dot\phi^2}{2},\label{sr2}\\
&&\ddot\phi+3H_E\dot\phi=0.\label{sr3}
\eea
In the plateau, the inflaton repeatedly oscillates between the left and right walls of the potential. We separately consider the time intervals dependent on the direction of the motion of the inflaton and derive the analytic solution in each regime.

The inflaton is reflected at the left side of the plateau at $\phi=\sqrt{6}\log\gamma$. We regard the reflection occurs instantly and define the first reflection at $t=t_{E1}$. After that, the inflaton reaches the right wall at $\phi=0$ and is reflected at $t=t_{E2}$. Thus, we can periodically define $t_{En}$ until the inflaton stops somewhere. 

First, let us consider the interval $t_{E0}<t_E<t_{E1}$.
Eliminating $\dot\phi$ from Eq.~\eqref{sr1} and \eqref{sr2} and then solving it, we obtain the Hubble parameter and the scale factor
\bea
H_E(t_E)&=&\f{H_{E0}}{3H_{E0}(t_E-t_{E0})+1}, \label{HE1} \\
a_E(t_E)&=&a_{E0}\kk{3H_{E0}(t_E-t_{E0})+1}^{1/3}. \label{aE1}
\eea
From Eq.~\eqref{sr1}, we obtain $\dot\phi=\pm \sqrt{6}H_E$. For $t_{E0}<t_E<t_{E1}$, since the inflaton moves in the left direction, we choose $\dot\phi=-\sqrt{6} H_E$. Therefore, the solution is given by
\be \phi(t_E)=-\sqrt{\f{2}{3}}\log\kk{3H_{E0}(t_E-t_{E0})+1}. \label{pE1} \ee

Next, we move to the interval $t_{E1}<t_E<t_{E2}$. By using $\dot\phi=+\sqrt{6}H_E$, we can derive
\bea
H_E(t_E)&=&\f{H_{E1}}{3H_{E1}(t_E-t_{E1})+1}, \label{HE2} \\ 
a_E(t_E)&=&a_{E1}\kk{3H_{E1}(t_E-t_{E1})+1}^{1/3}, \label{aE2} \\
\phi(t_E)&=&\sqrt{\f{2}{3}}\log\kk{3H_{E1}(t_E-t_{E1})+1}+\sqrt{6}\log\gamma \;. \label{pE2}
\eea
The matching conditions between the two intervals are determined by setting $\phi(t_{E1})=\sqrt{6}\log\gamma$ in the solution for $t_{E0}<t_E<t_{E1}$ as
\bea
t_{E1}&=&t_{E0}+\f{\gamma^{-3}-1}{3H_{E0}}, \\
H_{E1}&=&H_{E0}\gamma^{3},\\
a_{E1}&=&a_{E0}\gamma^{-1}.
\eea
By substituting them into the solutions \eqref{HE2} -- \eqref{pE2}, we obtain the same expressions as in Eqs.~\eqref{HE1} and \eqref{aE1} for $H_E(t_E)$ and $a_E(t_E)$. However, $\phi(t_E)$ is different from Eq.~\eqref{pE1}. It is given by
\be \phi(t_E)=\sqrt{\f{2}{3}}\log\kk{3H_{E0}(t_E-t_{E0})+1}+2\sqrt{6}\log\gamma. \ee

Likewise, by solving the recurrence equations, we conclude that the boundary conditions for general $n$ are
\begin{align}
t_{En}&=t_{E0}+\f{\gamma^{-3n}-1}{3H_{E0}}, \label{tEn} \\
H_{En}&=H_{E0}\gamma^{3n},
\label{eq9} \\
a_{En}&=a_{E0}\gamma^{-n},\\
\phi_{En}&=\left\{
\begin{array}{ll}
\displaystyle 0 &\quad (n:{\rm even})  \\
\displaystyle \sqrt{6}\log\gamma &\quad (n:{\rm odd}) 
\end{array}
\right. \;.
\end{align}
Especially, it is noteworthy that the following relation holds regardless of the parity of $n$:
\be 3H_{En}(t_{E}-t_{En})+1=\gamma^{3(n-1)}[3H_{E0}(t_{E}-t_{E0})+1] \;. \ee
Thus, the solutions for $t_{En-1}<t_E<t_{En}$ are
\bea
H_E(t_E)&=&\f{H_{E0}}{3H_{E0}(t_E-t_{E0})+1}, 
\label{eq5} \\
a_E(t_E)&=&a_{E0}\kk{3H_{E0}(t_E-t_{E0})+1}^{1/3}, 
\label{eq6} \\
\phi(t_E)&=&\left\{
\begin{array}{ll}
\displaystyle -\sqrt{\f{2}{3}}\log\kk{3H_{E0}(t_E-t_{E0})+1}-(n-1)\sqrt{6}\log\gamma &\quad (n:{\rm odd}) \\
\displaystyle \sqrt{\f{2}{3}}\log\kk{3H_{E0}(t_E-t_{E0})+1}+n\sqrt{6}\log\gamma &\quad (n:{\rm even})
\end{array}
\right. \;, \\
\dot\phi(t_E)&=&\left\{
\begin{array}{ll}
\displaystyle -\sqrt{6}H_E(t_E) &\quad (n:{\rm odd}) \\
\displaystyle \sqrt{6}H_E(t_E) &\quad (n:{\rm even}) 
\end{array}
\right. \;. \label{dphifrn}
\eea
Notice that the solutions for $\phi$ and $\dot \phi$ have different expressions for the different parity of $n$ because of its direction of the motion.

Once the solutions in the Einstein frame are at hand, it is straightforward to convert them to those in the Jordan frame. The Jordan frame time $t_J$ for $t_{En-1}<t_E<t_{En}$ is obtained by integrating $e^{-\phi/\sqrt{6}}$,
\be 
t_J(t_E)=
\left\{
\begin{array}{ll}
\displaystyle t_{Jn-1}+\f{\gamma^{n-1}}{4H_{E0}} \left[ \left\{ 3 H_{E0}(t_E-t_{E0})+1\right\}^{4/3}-\gamma^{-4(n-1)} \right] &\quad (n:{\rm odd}) \\
\displaystyle t_{Jn-1}+\f{\gamma^{-n}}{2H_{E0}} \left[ \left\{ 3 H_{E0}(t_E-t_{E0})+1\right\}^{2/3}-\gamma^{-2(n-1)} \right] &\quad (n:{\rm even})
\end{array}
\right. \;, \label{tJtEfr}
\ee
where $t_{Jn}$ is given by
\be
t_{Jn}=
\left\{
\begin{array}{ll}
\displaystyle t_{J0}+\f{(\gamma^4+\gamma^2+2)(\gamma^{-3(n-1)}-1)}{4H_{E0}(\gamma^4+\gamma^2+1)}+\f{\gamma^{-4}-1}{4H_{E0}\gamma^{3(n-1)}} &\quad (n:{\rm odd}) \\
\displaystyle t_{J0}+\f{(\gamma^4+\gamma^2+2)(\gamma^{-3n}-1)}{4H_{E0}(\gamma^4+\gamma^2+1)} &\quad (n:{\rm even})
\end{array}
\right. \;.
\label{eq4}
\ee

From Eqs.~\eqref{ta}, \eqref{HJ}, \eqref{eq5}, \eqref{eq6}, and \eqref{tJtEfr}, the Hubble parameter and the scale factor in the Jordan frame evolve as
\bea
H_J(t_J)&=&
\left\{
\begin{array}{ll}
\displaystyle \f{2\gamma^{3(n-1)}H_{E0}}{4\gamma^{3(n-1)}H_{E0}(t_J-t_{Jn-1})+1} &\quad (n:{\rm odd}) \\
\displaystyle 0 &\quad (n:{\rm even})
\end{array}
\right. \;, \label{HJfr} \\
a_J(t_J)&=&
\left\{
\begin{array}{ll}
\displaystyle a_{J0}\gamma^{-(n-1)}\kk{4\gamma^{3(n-1)}H_{E0}(t_J-t_{Jn-1})+1}^{1/2} &\quad (n:{\rm odd}) \\
\displaystyle a_{J0}\gamma^{-n} &\quad (n:{\rm even})
\end{array}
\right. \;. \\
\eea
$H_J(t_J)$ periodically takes discrete behaviors at $t_J=t_{Jn}$ and vanishes for even $n$. However, to be precise, $H_J$ is not exactly zero because we here neglect the contribution from the energy density of created radiation and the potential. We shall numerically confirm this point in the next section.
We define $H_{Jn}$ by $\lim_{\epsilon\to\pm 0} H(t_{Jn}+\epsilon)$ where $\pm$ for even and odd $n$, respectively. From the Hubble parameter \eqref{HJfr}, it is given by
\be H_{Jn}=2H_{En} = \gamma^{3n} H_{J0}. \ee
Here we used Eqs.~(\ref{eq8}) and (\ref{eq9}). 

Let us remark the time averaged behavior during the oscillation. Since Eq.~\eqref{tEn} implies $\log (t_{En}-t_{E0}) \propto n$, the time intervals when inflaton goes to the left and the right are equal in terms of the logarithmic Einstein-frame time. On the other hand, Eq.~\eqref{tJtEfr} yields $\log (t_J-t_{J0})\simeq p\log (t_E-t_{E0})$ with $p=4/3$ and $2/3$ for the left-directed regime and the right-directed regime, respectively. Hence, the left-directed regime is twice as long as the right-directed regime in terms of the logarithmic Jordan-frame time. Taking care of these facts, we can derive the average power of the Jordan frame quantities. For instance, as for time duration, since the average power is $(4/3+2/3)/(1+1)=1$, we can say $t_E$ is proportional to $t_J$ in average, {\it i.e.}, $\langle t_J-t_{J0}\rangle\propto t_E-t_{E0}$. Since $\log a_J\simeq q \log t_J$ with $q=1/2$ and $0$ for the left-directed regime and the right-directed regime respectively, the averaged power is $(1/2\times 2+0\times 1)/(2+1)=1/3$, namely, $\langle a_J(t_J)\rangle \propto (t_J-t_{J0})^{1/3}$.
Hence, the averaged Hubble parameter is written as
\begin{equation}
\langle H_J(t_J) \rangle = \frac{H_{J0}}{3 H_{J0} (t_J-t_{J0}) +1} 
\label{eq7} \;.
\end{equation}
Integrating this gives the averaged scale factor
\begin{equation}
\langle a_J(t_J) \rangle = a_{J0} \left[ 3 H_{J0} (t_J-t_{J0}) + 1 \right]^{1/3} \;.
\label{eq14}
\end{equation}

In summary, neglecting the small structure of the plateau and the energy density of the radiation enables us to solve the differential equations analytically. However, in the final phase of the reheating, they become important. We shall consider this effect in the next section.

\section{Numerical Results}%%%%%%%%%%%%%%%%%%%%%%%%%%%%%%%%%%%%%%%%%
\label{sec-nr}

As we derived in Sec.~\ref{sec-be}, the typical height of the local maximum of the potential plateau is $V_\max\propto \d$. The oscillation regime ends when the kinetic energy of scalaron is of the same order of $V_\max$. Thus, $\delta$ determines the end time of the reheating. However, since $\delta$ is the ratio between the energy scales of DE and of inflation, it has to be dramatically tiny value $\sim 10^{-120}$. Therefore, it is difficult to carry out numerical calculation for realistic model parameters. In this section, using the modest value of $\d$, we confirm the validity of the analytic solutions obtained in the previous section by comparing them with numerical results. Then we extrapolate the analytic solutions to the end of the reheating and estimate the reheating temperature as a function of the model parameters. Note that hereafter we explicitly fix the hat to normalized dimensionless physical variables, which was abbreviated in the previous section.

\subsection{Comparison with analytic solutions}%%%%%%%%%%%%%%%%%%%%%%%%%%%%%%%%%%%%%%%%%
\label{ssec-cwa}

%==================== Figure ====================
\begin{figure}[t]
\centering
\includegraphics[width=85mm]{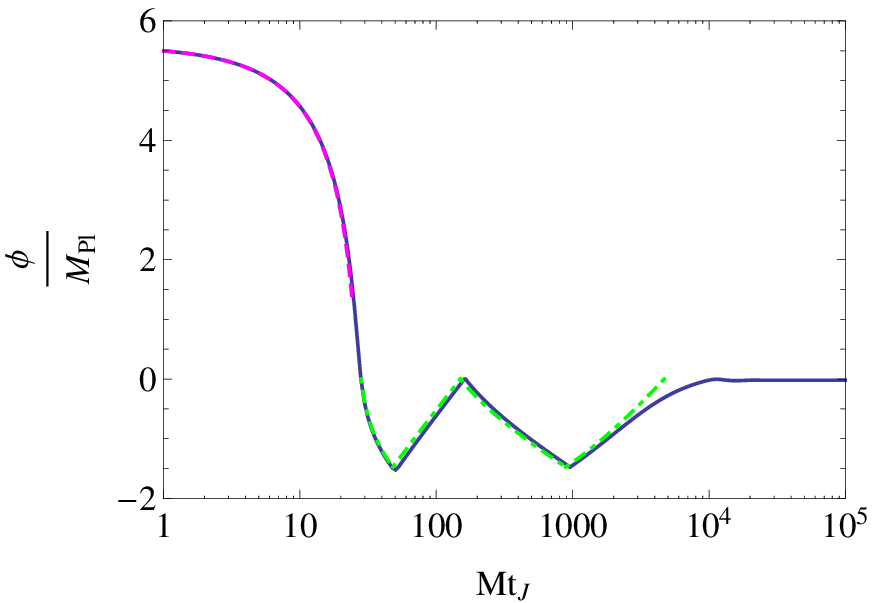}
\includegraphics[width=85mm]{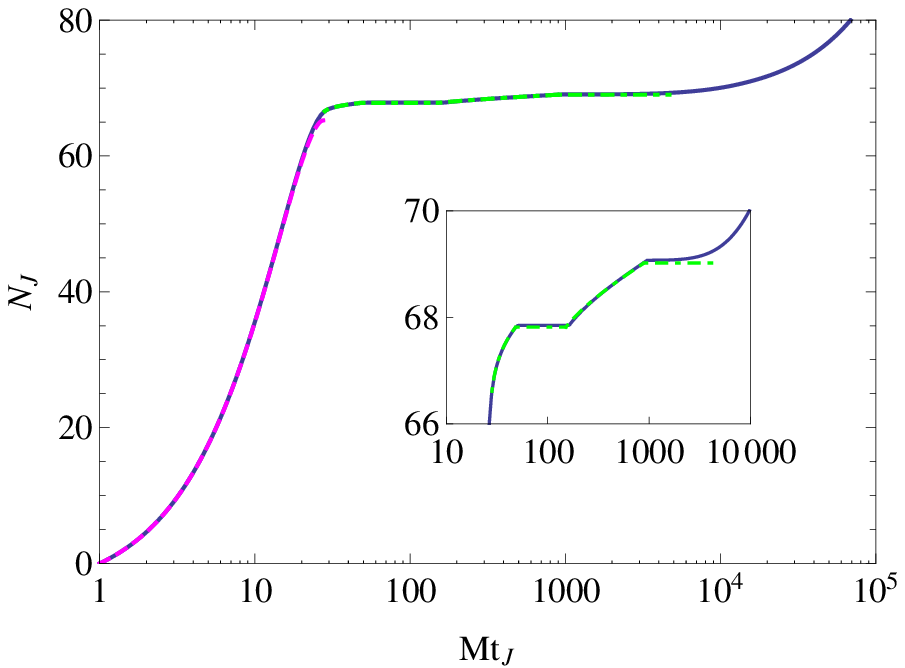}
\includegraphics[width=85mm]{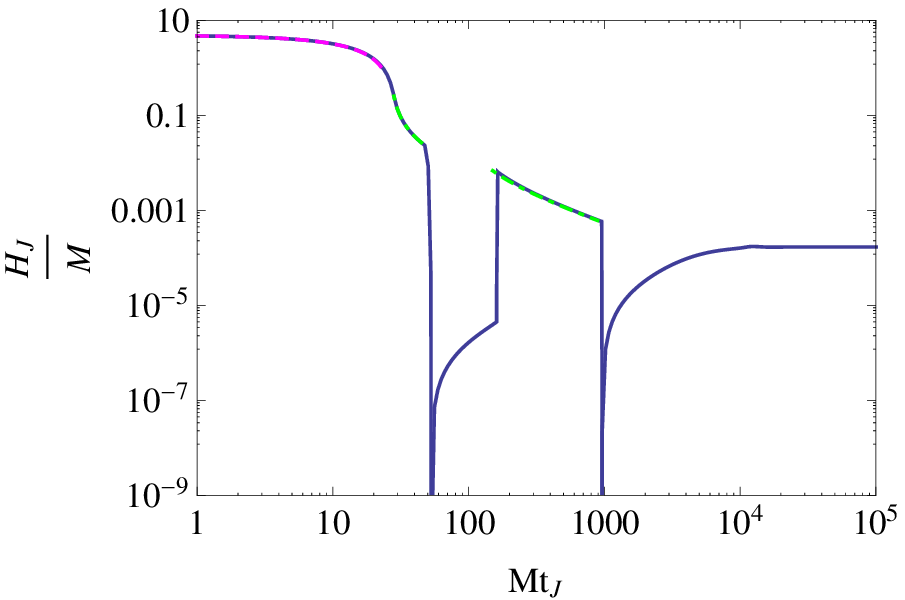}
\includegraphics[width=85mm]{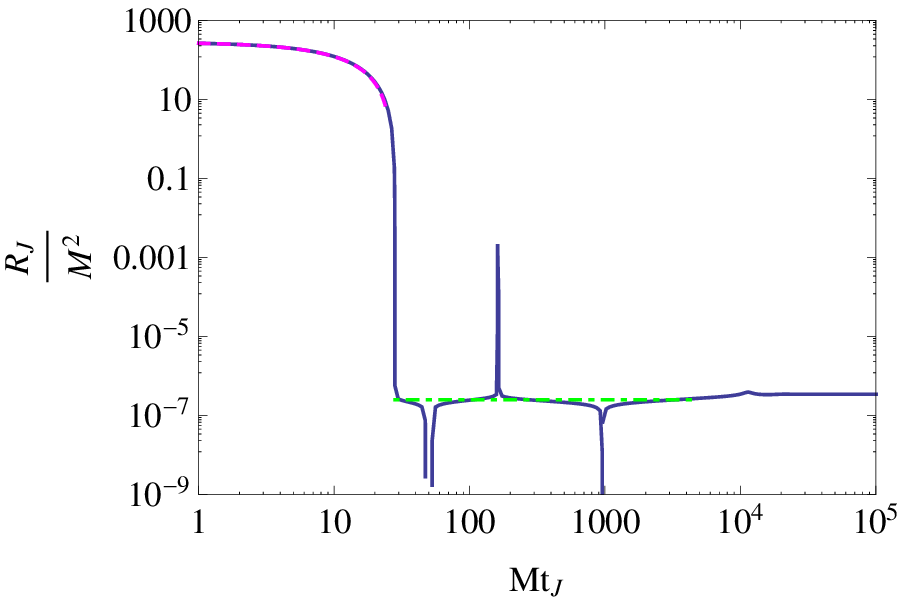}
\caption{Evolution of inflaton, $e$-folding number, Hubble parameter and Ricci scalar for model parameter $g=0.35, b=5, \d=5\times 10^{-8}, M/\Mpl=1.2\times 10^{-5}$. Numerical results (blue, solid), slow roll analytic solution (magenta, dashed) and fast roll analytic solution (green, dot-dashed) are presented.}
\label{fig:evol}
\end{figure}
%==================== Figure ====================

We performed numerical calculation in the Einstein frame, {\it i.e.}, we solved the coupled evolution equations: Eqs.~\eqref{Eeq2}, \eqref{Eeq3}, and \eqref{Jeq1}. In particular, to avoid the complexity due to nonminimal couplings in radiation and matter sectors, we use Eq.~\eqref{Jeq1}, instead of the continuous equation in the Einstein frame, by translating it to the Einstein frame with Eqs.~\eqref{FRJ}, \eqref{ta}, and \eqref{HJ}.  After the numerical calculation, we obtain physical quantities in the Jordan frame by the inverse conformal transformations.

Figure~\ref{fig:evol} illustrates the numerical results and the analytic solutions for a model parameter set: $g=0.35, b=5, \d=5\times 10^{-8}, M/\Mpl=1.2\times 10^{-5}$. Since observationally required value for $\d$ is too tiny to perform the numerical calculation, we chose this $\delta$ just for demonstration and compared its results with the analytic solutions. As mentioned before, the slow-roll approximation does not hold at the intermediate stage from the slow roll to the fast roll. So we used the boundary conditions obtained by the numerical calculation as the initial conditions for the fast-roll approximated solutions: $c_0 = 0.72$ and $\hat{H}_{J0} =0.26$ (These are different from the values obtained from slow-roll analytic solutions by about a factor of 2). The analytic solutions well approximate the numerical results. In the slow-roll regime, the scale factor undergoes quasi-de Sitter expansion. In the fast-roll regime, the inflaton oscillates two times inside the potential plateau. We see that the scale factor evolves as $a_J(t_J)\propto (t_J-t_{J0})^{1/2}$ and $a_J(t_J)\simeq \text{const}$ in order. Ricci scalar stays constant value $\simeq b\d$ except for several spikes during the reheating. In contrast to the analytical solution, $H_J$ does not vanish when the inflaton moves to the right direction in the potential plateau because of the contribution from the energy density $\rho_r$ of the created radiation. 
At late time, the fast-roll approximation becomes worse as the inflaton loses its kinetic energy. Finally, the inflaton reaches a false vacuum at $\phi=0$.

\subsection{End of the reheating}%%%%%%%%%%%%%%%%%%%%%%%%%%%%%%%%%%%%%%%%%
\label{ssec-er}

Since the above calculation have performed for $\d=5\times 10^{-8}$, the scalaron ends the oscillation quickly before radiation dominates the Universe. However, the scenario is different for $\d\sim 10^{-120}$: the reheating appropriately ends by radiation domination. This is because $V(0)$ and $V_{\rm{max}}$ is too small to compete with the inflaton kinetic energy. For the following, by using the analytic solutions, we estimate both the times when the scalaron stops the oscillation and when radiation dominates the Universe.

First, let us estimate when the oscillation stops. It is estimated from the equality time of the kinetic energy of the scalaron and the local maximum of the plateau: $\dot\phi^2/2\sim V_\max$. We use the analytic solution for $\dot\phi$ in Eq.~\eqref{dphifrn} and the asymptotic behavior $\langle \hat{H}_E \rangle \sim 1/3(\hat{t}_E-\hat{t}_{E0})\sim 1/3(\hat{t}_J-\hat{t}_{J0})$. Thus, the scalaron ceases the oscillation at 
\be \hat{t}_{Js}-\hat{t}_{J0} =\sqrt{\f{8(1-2g)}{3b\d}} , \ee
where we used $V_\max$ by Eq.~\eqref{V0Vmax}. Substituting $\d\sim 10^{-120}$, $M=1.2\times 10^{-5}\Mpl$, and $g,b \sim {\cal{O}}(1)$ yields $t_{Js}\sim 10^{46}~\text{GeV}^{-1}$, which is close to the Hubble time $H_0^{-1}$. To be precise, we should take the radiation and matter dominated epochs into account in the following evolution of the Universe, but it does not change the conclusion that the scalaron oscillation continues for the order of the Hubble time.

Next, let us estimate when the radiation dominates the Universe and the reheating ends. We compare the energy density of radiation due to particle creation and that of gravity. Since $R_J\simeq b\d$ during the oscillation in the plateau of the potential, we can expand $F$ around $R_J= b\d$ and take its linear order only,
\be F\simeq 1-g+\f{b\d}{3}+\mk{\f{1}{3}+\f{g}{\d}}(\hat{R}_J-b\d)\equiv e^{\sqrt{\f{2}{3}}\hat{\phi}}. \ee 
Thus, we can explicitly connect the Ricci scalar in the Jordan frame with the inflaton as
\be \hat{R}_J(t_E)=\f{3\d(e^{\sqrt{\f{2}{3}}\hat{\phi}(t_E)}-1+g(b+1))}{3g +\d}. \ee 
As we mentioned at the beginning of Sec.~\ref{ssec-fr}, the particle creation during the plateau oscillation phase is negligible because $\hat{R}_J\simeq b\delta \ll 1$. Therefore, $\rho_r$ is approximately given by
\be \langle \rho_r(t_J) \rangle =\rho_{r0}\mk{\f{\langle a_J(t_J) \rangle }{a_{J0}}}^{-4} \approx \frac{\rho_{r0}}{\left[ 3 H_{J0} (t_J-t_{J0})\right]^{4/3}}, \label{eq99} \ee
where we used Eq.~(\ref{eq14}). 
As for gravitational contribution, it is convenient to define the effective energy density of gravity by the equation of motion in the Jordan frame in Eq.~(\ref{eq11}):
\begin{align}
H_J^2 &= \frac{1}{3M_{\rm{Pl}}^2} (\rho_r+\rho_g) \;, 
\label{eq45} \\
\rho_g &\equiv \frac{3M_{\rm{Pl}}^2}{M^2} \left( g M^2 H_J^2 -2 H_J^{''} H_J +(H_J^{'})^2-6 H_J^{'} H_J^2 +g M^2 (H_J^{'}+H_J^2) \tanh \left[ \frac{R_J}{M^2 \delta}-b \right] \right. \nonumber \\
&\left. -\frac{g}{6} M^4\delta \log \left[ \frac{\cosh (R_J/M^2\delta -b)}{\cosh b} \right] - \frac{6g (H_J^{''} H_J + 4 H_J^{'} H_J^2)}{\delta \cosh^2 (R_J/M^2\delta-b)} \right) \;.
\label{eq42}
\end{align} 
When $\rho_r$ is negligible compared to $\rho_g$, from Eq.~(\ref{eq7}), the energy density of gravity is reduced to
\begin{equation}
\langle \hat{\rho}_g \rangle \approx 3 \langle \hat{H}_J \rangle ^2 \approx \frac{1}{3(\hat{t}_J-\hat{t}_{J0})^2}\;.
\label{eq44}
\end{equation}

%==================== Figure ====================
\begin{figure}[t]
\centering
\includegraphics[width=85mm]{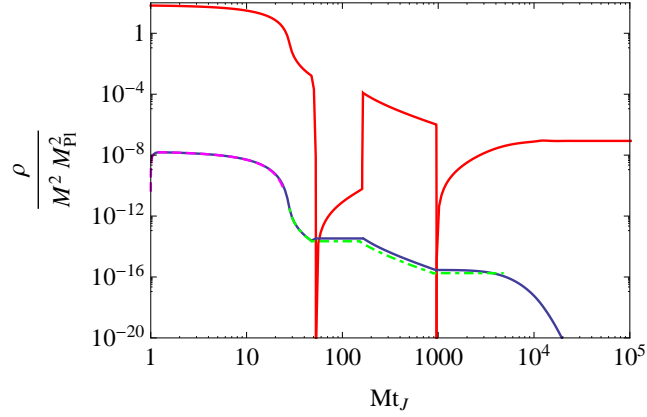}
\caption{Evolution of energy density of radiation $\hat \rho_r$ (blue) and effective energy density of gravity $\hat \rho_g\approx 3\hat H_J^2$ (red) for model parameter $g=0.35, b=5, \d=5\times 10^{-8}, M/\Mpl=1.2\times 10^{-5}$. For energy density of radiation, the analytic solutions for slow roll regime (magenta, dashed) and fast roll regime (green, dot-dashed) are also presented.}
\label{fig:rho}
\end{figure}
%==================== Figure ====================

Figure~\ref{fig:rho} represents the evolution of $\hat \rho_r$ and $\hat \rho_g\approx 3 \hat H_J^2$. Throughout the inflation and reheating, the energy density of the radiation is subdominant for our choice of $\d=5\times 10^{-8}$. The radiation is mainly created at the end of inflation and decays as $\rho_r\propto a^{-4}$. Since the scale factor evolves as $t_J^{1/2}$ and constant periodically, $\rho_r$ correspondingly evolves as $t_J^{-2}$ and constant during the left-directed and right-directed regimes, respectively. As we mentioned, $3 \hat H_J^2$ does not vanish because of the contribution from the radiation. At $Mt_J\simeq 50$ and $1000$, $3 \hat H_J^2$ is smaller than $\rho_r$. This is because the required step width at the abrupt transition is so tiny that $\hat H_J$ over-decreases. This tiny discrepancy is unimportant and does not affect the subsequent evolution.

The end time of the reheating and the reheating temperature $T_{\rm{reh}}$ is defined by the condition $\langle \rho_r \rangle =\langle \rho_g \rangle$. Using Eqs.~(\ref{eq99}) and (\ref{eq44}), we obtain
\begin{equation}
\hat{t}_{J,\rm{reh}} - \hat{t}_{J0} \approx \frac{\sqrt{3} \hat{H}_{J0}^2}{\hat{\rho}_{r0}^{3/2}}
 \approx 3.8 \times 10^5 (c_0 g_*)^{-3/2} \hat{H}_{J0}^2 \left( \frac{M}{M_{\rm{Pl}}}
\right)^{-3} \;,
\label{eq54}
\end{equation}
and 
\begin{equation}
\frac{T_{\rm{reh}}}{M} = \hat{H}_{J0} \left( \frac{a_{J0}}{\langle a_{J,\rm{reh}} \rangle} \right) \approx \sqrt{\frac{\hat{\rho}_{r0}}{3}} \approx 9.6 \times 10^{-3} (c_0 g_*)^{1/2} \left( \frac{M}{M_{\rm{Pl}}} \right) \;.
\label{eq12}
\end{equation}
Since these equations include $H_{J0}$ and $\rho_{r0}$, the end of the reheating is sensitive to the boundary conditions at the transition from the slow-roll regime to the fast-roll regime. For $g_*=106.75$, $M/\Mpl=1.2\times 10^{-5}$, and numerically determined values, $c_0=0.72$, $\hat{H}_{J0}\approx 0.26$, the time and temperature of the reheating are $\hat{t}_{J,\rm{reh}} - \hat{t}_{J0} \approx 2.2\times 10^{16}$ and $T_{\rm{reh}} \approx 3.0 \times 10^7\,{\rm{GeV}}$, respectively.

\section{Constraints on the model parameters}%%%%%%%%%%%%%%%%%%%%%%%%%%%%%%%%%%%%%%%%%
\label{sec-rtcn}

In this section, based on our analytic solutions and numerical results obtained in the previous sections, we discuss the allowed ranges of the model parameters. First, we present the constraint on the energy scale $M$ in Sec.~\ref{ssec-conm}. It is fixed from the amplitude normalization of the CMB. Second, we consider the other parameters, $g$, $b$, and $\d$ in Sec.~\ref{ssec-congbd}. Once $M$ is determined, we can predict the other observable of the inflation and discuss its consistency with observations. $g$, $b$, and $\delta$, cannot be constrained from observational data because the reheating temperature in Eq.~(\ref{eq12}) is independent of those parameters. However, since the $gR^2$-AB model must realize current cosmic acceleration, its magnitude and stability constrain the allowed range of $g$, $b$, and $\delta$.

\subsection{Constraint on $M$}%%%%%%%%%%%%%%%%%%%%%%%%%%%%%%%%%%%%%%%%%
\label{ssec-conm}

When $R_J \gg M^2$, the $gR^2$-AB model~(\ref{eq58}) can be approximated to $R^2$ inflation, in which the primordial spectrum of a scalar mode at the leading order in the slow-roll parameter $\epsilon_1$ is given by~\cite{Hwang:2001pu}
\begin{equation}
{\cal{P}}_S \approx \frac{1}{96 \pi^2 \epsilon_1^2} \left( \frac{M}{M_{\rm{Pl}}} \right)^2 \;, 
\label{eq53} 
\end{equation}
where $\epsilon_1 \equiv -H_J^{'}/H_J^2$. This slow-roll parameter is related to the $e$-folding number between the end of inflation and the horizon crossing of the mode whose comoving wave number $k$ corresponds to the CMB scale today.
From the analytic solutions during the slow-roll regime, Eqs.~(\ref{HJ}), (\ref{phisr}) - (\ref{HEsr}), the Hubble parameter in the Jordan frame is written as
\begin{equation}
\hat{H}_J(t_J) = \frac{3\, \tau(\hat{t}_J)-1}{6\,\tau^{1/2}(\hat{t}_J)} \;, \quad \quad \tau(\hat{t}_J) \equiv d_{\rm{ini}}^2 -\frac{2}{3} (\hat{t}_J-\hat{t}_{J,{\rm{ini}}}) \left[ d_{\rm{ini}} - \frac{1}{6} (\hat{t}_J-\hat{t}_{J,{\rm{ini}}}) \right]
\label{eq13} \;.
\end{equation}
For $d_{\rm{ini}} \gg 1$, expanding Eq.~(\ref{eq13}) in powers of $t_J$ around $t_{J,{\rm{ini}}}$ gives 
\begin{equation}
\hat{H}_J(\hat{t}_J) \approx \hat{H}_{J,{\rm{ini}}} -\frac{1}{6} (\hat{t}_J-\hat{t}_{J,{\rm{ini}}}) \;,
\end{equation}
where $\hat{H}_{J,{\rm{ini}}} = d_{\rm{ini}}/2$.
The $e$-folding number between the end of inflation at $t_{J,{\rm{end}}}$ and the horizon crossing of the CMB mode at $t_{Jk}$ is given by
\begin{equation}
N_k = \int_{t_{Jk}}^{t_{J,\rm{end}}} H_J dt_J \approx -\frac{H_{Jk}^2}{2H^{'}_{Jk}} = \frac{1}{2 \epsilon_1 (t_{Jk})} \;, 
\end{equation}
where $H_{Jk} \equiv H_J(t_k)$ and we used the fact that $H^{'}_{Jk}$ is constant when we performed the integration. Then Eq.~(\ref{eq53}) is expressed as
\begin{equation}
{\cal{P}}_S \approx \frac{N_k^2}{24\pi^2} \left( \frac{M}{M_{\rm{Pl}}} \right)^2 \;.
\end{equation}
Using Eq.~(\ref{eq99}), we obtain the $e$-folding number when the comoving scale of CMB crosses the horizon during inflation:
\begin{equation}
N_k \approx 66.2 -\frac{1}{2} \log \left( \frac{1-\Omega_m}{0.7} \right) -\frac{1}{4} \log \left[ \left( \frac{c_0}{0.72} \right) \left( \frac{g_{\ast}}{106.75} \right) \right] \;.
\end{equation}
Since the parameters $\Omega_m$ and $g_{\ast}$ hardly change $N_k$, we set it to $N_k=66$. From the temperature fluctuation of CMB anisotropy~\cite{Komatsu:2009}, the amplitude of the power spectrum, ${\cal{P}}_S =(2.445 \pm 0.096) \times 10^{-9}$ at $k=0.002\,{\rm{Mpc}}^{-1}$, fixes the parameter $M$ to
\begin{equation}
\frac{M}{M_{\rm{Pl}}} \approx 1.2 \times 10^{-5} \;. 
\end{equation}
At the CMB scale, the spectral indices of the scalar and tensor modes and the tensor-to-scalar ratio are given by~\cite{Hwang:2001pu} 
\begin{align}
n_S-1 &\equiv \left. \frac{d \log {\cal{P}}_S(k)}{d \log k} \right|_{k=aH}
\approx -\frac{2}{N_k} \;, \\
n_T &\equiv \left. \frac{d \log {\cal{P}}_T(k)}{d \log k} \right|_{k=aH} \approx -\frac{3}{2 N^2_k} \;, \\
r &\equiv \frac{{\cal{P}}_T}{{\cal{P}}_S} \approx \frac{12}{N_k^2} \;,
\end{align}
at the leading order in the slow-roll parameter. For the above choice of $N_k$, $n_S \approx 0.97$ and $r \approx 2.8 \times 10^{-3}$, which are consistent with observational bounds~\cite{Komatsu:2009}.

\subsection{Constraints on $g$, $b$, and $\delta$}%%%%%%%%%%%%%%%%%%%%%%%%%%%%%%%%%%%%%%%%%
\label{ssec-congbd}

The parameters $g$, $b$ and $\delta$ considerably alter the dynamics of the reheating in the $gR^2$-AB model. However, as seen from Eqs.~(\ref{eq99}) and (\ref{eq12}), the reheating temperature and the radiation energy density at that time does not depends on these parameters. So the constraint on $g$, $b$, and $\delta$ comes not from the CMB observation but from a stability condition of a de-Sitter vacuum.

\begin{figure}[t]
\begin{center}
\includegraphics[width=8.5cm]{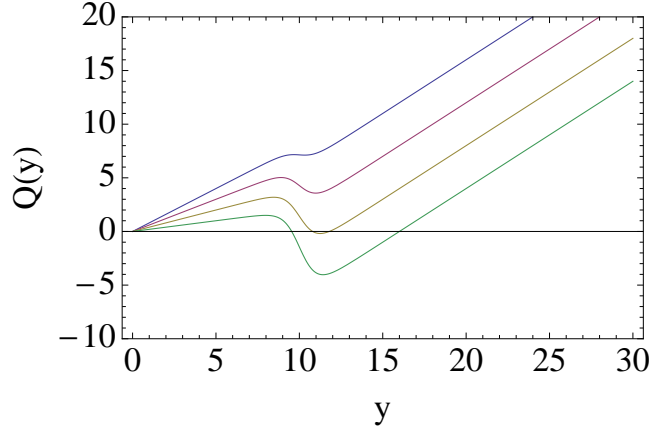}
\caption{Function $Q(y)$ for $b=10$ and $g=0.1$ (blue), $0.2$ (red), $0.3$ (yellow), and $0.4$ (green).}
\label{fig17}
\end{center}
\end{figure}

\begin{figure}[t]
\begin{center}
\includegraphics[width=8.5cm]{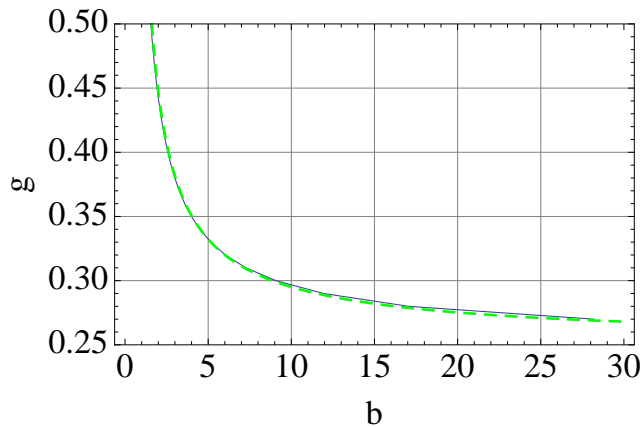}
\caption{Allowed parameter region of $g$ and $b$. Numerical solution (solid curve), fitting (dotted curve). Above this curve, stable de-Sitter solutions exist.}
\label{fig16}
\end{center}
\end{figure}

From Eq.~(\ref{eq10}), 
\begin{equation}
3 \Box F(R_J) +R_J F(R_J) -2 f(R_J) = 8\pi G T_J \;,
\end{equation}
where $T_J \equiv g_J^{\mu\nu} T^J_{\mu\nu}$ and $\Box \equiv (1/\sqrt{-g_J})\, \partial_{\mu} \left[ \sqrt{-g_J}\, \partial^{\mu} \right] $. For the existence of a stable solution of a de-Sitter vacuum ($R_J={\rm{const}}.$, $T_J=0$), the following equation has to be satisfied:
\begin{equation}
R_J F(R_J) -2 f(R_J)=0 \;.
\label{eq59}
\end{equation}
For $R_J \ll M^2$, substituting Eq.~(\ref{eq58}) into Eq.~(\ref{eq59}) and using $R_{\rm{vac}} \approx 4 g b M^2 \delta$ lead to the equation  
\begin{equation}
Q(y) \equiv y-4gb +2 g \left[ \log \left( 1+e^{-2(y-b)} \right) +\frac{y}{1+e^{2(y-b)}} \right] =0 \;. 
\end{equation}
where $y \equiv R_J/M^2\delta$. This function $Q(y)$ typically has the shape shown in Fig.~\ref{fig17}. Therefore, a stable de-Sitter vacuum exists if $Q^{'}(y_0)=0$ has the solution $y=y_0>1$ at which $Q^{''}(y_0)>0$ and $Q(y_0) \leq 0$. The boundary of the allowed parameter region of $b$ and $g$ can be obtained by solving $Q(y_0)=0$ and $Q^{'}(y_0)=0$ under the condition $Q^{''}(y_0)>0$. We cannot solve the above equations analytically. Instead, 
we fit the numerical solution and obtain the allowed region for $g$ as  
\begin{equation}
\frac{1}{4} + \frac{0.28}{(b-0.46)^{0.81}} \leq g \leq \frac{1}{2} \;. \label{gcon}
\end{equation}
This region is shown in Fig.~\ref{fig16}.

%==================== Figure ====================
%==================== Figure ====================

%==================== Figure ====================
%==================== Figure ====================

Once the parameters $M$, $g$, and $b$ are fixed, $\delta$ should be determined so that the current observation of accelerating expansion is reproduced. With the Ricci curvature of the present universe, $R_{\rm{vac}} \sim 10^{-120} M_{\rm{Pl}}^2$, the parameter $\delta$ is given by
\begin{equation}
\delta = \frac{R_{\rm{vac}}}{2g M^2 (b+\log [ 2\cosh b])} \approx \frac{1}{4g b} \frac{R_{\rm{vac}}}{M^2} \;.
\end{equation}

\section{Conclusions and discussion}%%%%%%%%%%%%%%%%%%%%%%%%%%%%%%%%%%%%%%%%%
\label{sec-cn}

We have studied the inflation and reheating dynamics in $f(R)$ gravity, especially in $gR^2$-AB model. This model is capable to describe both accelerated expansions in the early Universe and the present time. In the Einstein frame, the inflaton potential of this model possesses a plateau and a false vacuum in the bottom of the potential. These are different features from original $R^2$ inflation model and they significantly change the reheating process. We have derived the analytic solutions in the slow-roll inflation regime and the fast-roll oscillation (reheating) regime. We have also carried out the numerical computation including the backreaction from particle creation, and have confirmed that both results agree well. According to the existence of the potential plateau, the particle creation via gravitational reheating mainly occurs in the slow-roll regime and is inefficient during the the fast-roll oscillation regime. Consequently, in contrast to the $R^2$ inflationary scenario, the reheating era lasts longer. Another interesting feature of this model is that the averaged time evolution of a scale factor is proportional to $t_J^{1/3}$ because of the periodic abrupt changes of the Hubble parameter.

Based on these results obtained from our analytic and numerical calculations, we have given the constraints on the model parameters. The parameter $M$ is pinned down by the observational amplitude of CMB temperature fluctuations. Also the value of $\delta$ is selected to correctly reproduce the current accelerated expansion of the Universe. On the other hand, the parameters $g$ and $b$ are poorly constrained because these parameters affect only the reheating dynamics after the inflation. To more tightly constrain $g$ and $b$, we need observations that can probe at much smaller scales than those of CMB and large-scale galaxy surveys. In the future searches for primordial black holes and the direct detection experiments of gravitational waves would provide new observational windows for the reheating dynamics in modified gravity.

\begin{acknowledgments}%%%%%%%%%%%%%%%%%%%%%%%%%%%%%%%%%%%%%%%%%
We would like to thank A.~A.~Starobinsky, T.~Suyama and J.~Yokoyama for helpful discussions and valuable comments. This work was supported in part by JSPS Research Fellowships for Young Scientists (H.M.) and Grant-in-Aid for JSPS Fellows (A.N.).
\end{acknowledgments}

\end{document}